\documentclass[
    ,final            % use final for the camera ready runs
  ]
  {aipproc}

\layoutstyle{6x9}

\newcommand{\ba}{\begin{eqnarray}}
\newcommand{\ea}{\end{eqnarray}}

\begin{document}

\title{Baryons in the unquenched quark model}


\author{R. Bijker}
{address={Instituto de Ciencias Nucleares, Universidad Nacional Aut\'onoma de M\'exico, 
AP 70-543, 04510 Mexico DF, Mexico}}
\author{S. D{\'{\i}}az-G\'omez}
{address={Instituto de Ciencias Nucleares, Universidad Nacional Aut\'onoma de M\'exico, 
AP 70-543, 04510 Mexico DF, Mexico}}
\author{M.A. Lopez-Ruiz}
{address={Physics Department and Center for Exploration of Energy and Matter, Indiana University, 
Bloomington, IN 47408, USA}}
\author{E. Santopinto}
{address={Istituto Nazionale di Fisica Nucleare, Sezione di Genova, via Dodecaneso 33, I-16146 Italy}}

\begin{abstract}
In this contribution, we present the unquenched quark model as an extension of the 
constituent quark model that includes the effects of sea quarks via a $^{3}P_{0}$ 
quark-antiquark pair-creation mechanism. Particular attention is paid to the spin and 
flavor content of the proton, magnetic moments and $\beta$ decays of octet baryons.
\end{abstract}

\maketitle

\section{Introduction}

In the constituent quark model (CQM) hadrons are described as a system of 
constituent (or valence) quarks and antiquarks, $qqq$ for baryons and $q \bar{q}$ 
for mesons. Despite the success of the quark model, there is strong evidence for the 
existence of exotic degrees of freedom (other than valence quarks) in hadrons 
from CQM studies of electromagnetic and strong couplings of baryons that are, on 
average, underpredicted by CQMs \cite{CRreview}. More direct evidence for the 
importance of quark-antiquark components in the proton comes from measurements of the 
$\bar{d}/\bar{u}$ asymmetry in the nucleon sea \cite{Kumano,GarveyPeng} and the proton 
spin crisis \cite{protonspin}. 
The pion cloud in the nucleon holds the key to understand the flavor asymmetry and the 
spin-crisis of the proton \cite{Kumano,GarveyPeng,Speth}. Moreover, angular momentum 
conservation of the pionic fluctuations of the nucleon leads to a relation between the 
flavor asymmetry and the contribution of orbital angular momentum to the spin of the proton 
${\cal A}(p) = \Delta L$ \cite{Garvey}. 
 
The aim of this contribution is to discuss the role of valence and sea quarks in the nucleon 
in the framework of a simplified version of the unquenched quark model (UQM) in which 
only the effects of the pion cloud is taken into account. It is shown that the pion cloud 
offers a qualitative understanding of the results obtained in previous numerical studies 
\cite{uqm}, and thus provides important insights into the properties of the nucleon. 

\section{Unquenched quark model}

\begin{figure}[h]
\centering
\setlength{\unitlength}{1pt}
\begin{picture}(145,130)(0,-25)
\put( 20, 20) {\line(1, 0){50}}
\put( 20, 40) {\line(1, 0){50}}
\put( 20, 60) {\line(1, 0){50}}
\put( 70, 60) {\line(2, 1){50}}
\put( 70, 40) {\line(2,-1){50}}
\put( 70, 20) {\line(2,-1){50}}
\put( 90, 50) {\line(2, 1){30}}
\put( 90, 50) {\line(2,-1){30}}
\put( 20, 20) {\vector(1, 0){25}}
\put( 20, 40) {\vector(1, 0){25}}
\put( 20, 60) {\vector(1, 0){25}}
\put( 70, 60) {\vector(2, 1){40}}
\put( 70, 40) {\vector(2,-1){40}}
\put( 70, 20) {\vector(2,-1){40}}
\put(120, 65) {\vector(-2,-1){15}}
\put( 90, 50) {\vector(2,-1){20}}
\put(  5, 17) {$q_1$}
\put(  5, 37) {$q_2$}
\put(  5, 57) {$q_3$}
\put(125, -8) {$q_1$}
\put(125, 12) {$q_2$}
\put(125, 32) {$q$}
\put(125, 62) {$\bar{q}$}
\put(125, 82) {$q_3$}
\end{picture}
\caption{\small Schematic quark line diagram for $A \rightarrow B C$.}
\label{diagram}
\end{figure}
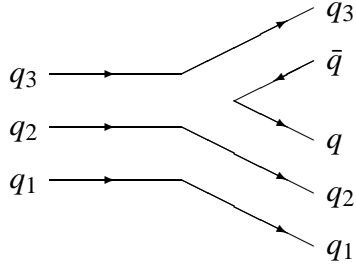

The unquenched quark model developed in \cite{uqm} is motivated by earlier 
studies on extensions of the quark model in which the $q\bar{q}$ pairs are created 
in the $^{3}P_0$ state with the quantum numbers of the vacuum \cite{tornqvist,baryons}. 
The present approach is based on a CQM to which the quark-antiquark pairs are added as 
a perturbation, employing a $^{3}P_0$ model for the $q \bar{q}$ pair creation. 
The pair-creation mechanism is inserted at the quark level and the one-loop diagrams 
are calculated by summing over the intermediate baryon-meson states (see Fig.~\ref{diagram}).

Under these assumptions, the baryon wave function consists of a zeroth order three-quark 
configuration $\mid A \rangle$ plus a sum over higher Fock components due to the creation 
of quark-antiquark pairs. The resulting baryon wave function is given by \cite{uqm} 
\ba 
\left| \psi_A \right> = {\cal N} \left[ \left| A \right>  
+ \sum_{BC l J} \int d \vec{K} k^2 dk \, \left| BC,l,J; \vec{K},k \right> \, 
\frac{ \left< BC,l,J; \vec{K},k \left| T^{\dagger} \right| A \right> } 
{\Delta E_{BC}(k)} \right] ~, 
\label{wf1}
\ea
where $\Delta E_{BC}(k) = M_A - E_B(k) - E_C(k)$ is the energy difference calculated in the rest 
frame of the initial baryon $A$. The operator $T^{\dagger}$ creates a quark-antiquark pair in the 
$^{3}P_0$ state with the quantum numbers of the vacuum: $L=S=1$ and $J=0$.  
The $^{3}P_{0}$ transition amplitude can be expressed as  \cite{roberts}
\ba  
\langle BC,l,J; \vec{K},k | T^{\dagger} | A \rangle 
= \delta(\vec{K}) \, M_{A \rightarrow BC}(k)
\ea 
where $\delta(\vec{K})$ is a consequence of momentum conservation in the rest frame of A. 

\subsection{Flavor and spin content}

In this contribution, we employ a simplified version of the UQM in which only the contribution  
of the pion cloud is taken into account. Table~\ref{spinflavor} shows the results for the flavor 
and spin content of the proton. In the UQM, the coefficients appearing in Table~\ref{spinflavor} 
are expressed in terms of integrals over the relative momentum $k$ 
\ba
a^2 &\rightarrow& \int k^2 dk \frac{|M_{N \rightarrow N \pi}(k)|^2}{\Delta E_{N \pi}^2(k)} ~,
\nonumber\\
b^2 &\rightarrow& \int k^2 dk \frac{|M_{N \rightarrow \Delta \pi}(k)|^2}{\Delta E_{\Delta \pi}^2(k)} ~,
\nonumber\\
2ab &\rightarrow& \int k^2 dk \frac{M^{\ast}_{N \rightarrow N \pi}(k) M_{N \rightarrow \Delta \pi}(k) 
+  M^{\ast}_{N \rightarrow \Delta \pi}(k) M_{N \rightarrow N \pi}(k)} 
{\Delta E_{N \pi}(k) \Delta E_{\Delta \pi}(k)} ~,
\ea
which only depend on the $^{3}P_{0}$ coupling strength. The results for the UQM in Table~\ref{spinflavor} 
are also valid for the meson-cloud model in which the coefficients $a$ and $b$ multiply the $N \pi$ and 
$\Delta \pi$ components of the nucleon wave function \cite{Garvey}. The $ab$ term denotes the contribution 
from the cross terms between the $N \pi$ and $\Delta \pi$ components. In the UQM, the value of the cross term 
$ab$ is not equal to the product of $a$ and $b$, although the numerical values are close. Since the UQM contains 
the full spin and isospin structure, it satisfies the relation between the flavor asymmetry and the 
contribution of the orbital angular momentum to the spin of the proton ${\cal A}(p)=\Delta L$ \cite{Garvey}, 
and therefore $\Delta \Sigma = 1-2\Delta L$. In the absence of the pion cloud ($a^2=b^2=2ab=0$) 
we recover the results of the CQM. 

\begin{table}
\centering
\caption{\small Spin and flavor content of the proton in he constituyent quark model (CQM) 
and the unquenched quark model (UQM), normalized to the flavor asymmetry using the E866/NuSea 
value \cite{Towell} (UQM1) and using the NMC value \cite{NMC} (UQM2).} 
\label{spinflavor}
\vspace{15pt}
\begin{tabular}{cccrrc}
\noalign{\smallskip}
\hline
\noalign{\smallskip}
& CQM & UQM & UQM1 & UQM2 & Exp \\
\noalign{\smallskip}
\hline
\noalign{\smallskip}
${\cal A}(p)=\Delta L$ & $0$ & $\frac{2a^2-b^2}{3(1+a^2+b^2)}$ 
& $*0.118$ & $*0.158$ & $ 0.118 \pm 0.012$ \\
&&&&& $0.158 \pm 0.010$ \\
\noalign{\smallskip}
$\Delta u$ & $ \frac{4}{3}$ & $ \frac{4}{3}-\frac{38a^2+b^2-16ab\sqrt{2}}{27(1+a^2+b^2)}$ 
& $ 1.132$ & $ 1.064$ & $ 0.842 \pm 0.013$ \\ 
\noalign{\smallskip}
$\Delta d$ & $-\frac{1}{3}$ & $-\frac{1}{3}+\frac{2a^2+19b^2-16ab\sqrt{2}}{27(1+a^2+b^2)}$ 
& $-0.368$ & $-0.380$ & $-0.427 \pm 0.013$ \\
\noalign{\smallskip}
$\Delta s$ & $0$ & $ 0$ & $ 0.000$ & $ 0.000$ & $-0.085 \pm 0.018$ \\
\noalign{\smallskip}
$\Delta \Sigma = \Delta u + \Delta d + \Delta s$ & $1$ & $1-\frac{4a^2-2b^2}{3(1+a^2+b^2)}$ 
& $ 0.764$ & $ 0.684$ & $ 0.330 \pm 0.039$ \\
\noalign{\smallskip}
$g_A = \Delta u - \Delta d$ & $\frac{5}{3}$ & $\frac{5}{3}-\frac{40a^2+20b^2-32ab\sqrt{2}}{27(1+a^2+b^2)}$ 
& $ 1.500$ & $ 1.444$ & $1.2723 \pm 0.0023$ \\
\noalign{\smallskip}
\hline
\end{tabular}
\end{table}

The results for the spin and flavor content of the proton are normalized to the proton flavor 
asymmetry. The fourth column is normalized to the E866/NuSea value \cite{Towell}, and the fifth 
column to the somewhat higher NMC value \cite{NMC}. The experimental values of the spin content 
were obtained by the HERMES \cite{Hermes} and the COMPASS \cite{Compass} Collaborations. 
Table~\ref{spinflavor} shows the results from the HERMES Collaboration.   

\subsection{Magnetic moments}

The magnetic moments of the octet baryons constitute one of the early successes of the 
constituent quark model. Hence, for any extension of the quark model, it is important to verify 
whether the good agreement of the CQM is maintained. Table~\ref{moments} shows that this is indeed 
the case for the unquenched quark model. Just as for the CQM, the quark magnetic moments 
are fitted to the magnetic moments of the proton, neutron and $\Lambda$ hyperon.  

\begin{table}[ht]
\centering
\caption{\small Magnetic moments. The experimental values are taken from Ref.~\cite{PDG}.}
\label{moments}
\vspace{15pt}
\begin{tabular}{crrrc}
\noalign{\smallskip}
\hline
\noalign{\smallskip}
& CQM & UQM1 & UQM2 & Exp \\
\noalign{\smallskip}
\hline
\noalign{\smallskip}
$p$       & $ 2.793$ & $ 2.793$ & $ 2.793$ & $ 2.793$ \\
$n$       & $-1.913$ & $-1.913$ & $-1.913$ & $-1.913$ \\
$\Lambda$ & $-0.613$ & $-0.613$ & $-0.613$ & $-0.613 \pm 0.004$ \\
\noalign{\smallskip}
\hline
\noalign{\smallskip}
$\Sigma^+$         & $ 2.673$ & $ 2.589$ & $ 2.509$ & $ 2.458 \pm 0.010$ \\
$\Sigma^0$         & $ 0.791$ & $ 0.783$ & $ 0.751$ & \\
$\Sigma^-$         & $-1.091$ & $-1.023$ & $-1.007$ & $-1.160 \pm 0.025$ \\
$\Xi^0$            & $-1.435$ & $-1.359$ & $-1.290$ & $-1.250 \pm 0.014$ \\
$\Xi^-$            & $-0.493$ & $-0.530$ & $-0.552$ & $-0.651 \pm 0.003$ \\
$\Sigma^0/\Lambda$ & $ 1.630$ & $ 1.640$ & $-1.638$ & $ 1.61  \pm 0.08 $ \\
\noalign{\smallskip}
\hline
\end{tabular}
\end{table}

\subsection{Fluctuations}

In the UQM, it is straightforward to calculate the fluctuation probabilities. 
The probability that a proton fluctuates in $n \pi^+$ is given by
\ba
\left| \left< n \pi^+ | p \right> \right|^2 \;=\; \frac{2a^2}{3(1+a^2+b^2)} ~,
\ea
whereas the total probability for a pion fluctuation of the proton is given by   
\ba
\left| \left< N \pi | p \right> \right|^2 + \left| \left< \Delta \pi | p \right> \right|^2 
\;=\; \frac{a^2+b^2}{1+a^2+b^2} ~.
\ea
The numerical results are shown in Table~\ref{fluctuations}. The UQM1 values are in good  
agreement with the experimental values as determined in an analysis of forward neutron 
production in electron-proton collisions by the H1 and ZEUS Collaborations at DESY 
\cite{Povh,Rosina1}, and in a study of the quark distribution functions measured in 
Drell-Yan experiments and semi-inclusive DIS experiments \cite{Chang}.  
The UQM2 values are about 30 \% higher than the UQM1 values.  

\begin{table}[ht]
\centering
\caption{\small Pion fluctuations of the proton.}
\label{fluctuations}
\vspace{15pt}
\begin{tabular}{crrcc}
\noalign{\smallskip}
\hline
\noalign{\smallskip}
& UQM1 & UQM2 & Exp & Ref \\
\noalign{\smallskip}
\hline
\noalign{\smallskip}
$\left| \left< n \pi^+ | p \right> \right|^2$ 
& 0.180 & 0.241 & $0.17 \pm 0.01$ & \cite{Povh,Rosina1} \\
$\left| \left< N \pi | p \right> \right|^2 + \left| \left< \Delta \pi | p \right> \right|^2$ 
& 0.455 & 0.609 & 0.470 & \cite{Chang} \\
\noalign{\smallskip}
\hline
\end{tabular}
\end{table}

\section{Summary and conclusions}

In this contribution, we studied the properties of the proton in the framework of the unquenched quark 
model in which the $^{3}P_0$ coupling strength was normalized to the observed value of the proton 
flavor asymmetry. It was shown that whereas the pion fluctuations maintain the good results of the 
constituent quark model for the magnetic moments, they help to understand the discrepancies between the 
CQM and the experimental data. Their inclusion leads to a reduction of quark model value of $\Delta u$ 
and $g_A$, and give rise to a sizeable contribution (25 - 30 \%) of orbital angular momentum to the spin 
of the proton. In addition, it was found that the probabilities for pion fluctuations in the UQM are in 
good agreement with the values determined in analyses of the available experimental data. 

\begin{theacknowledgments}

This work was supported in part by grant IN107314 from DGAPA-PAPIIT, Mexico.

\end{theacknowledgments}

\bibliographystyle{aipproc}

\end{document}